\documentclass[useAMS,usenatbib,usenatbib]{mn2e}    
\usepackage{amssymb}  
\usepackage{graphicx}
\def\be{\begin{equation}} 
\def\en{\end{equation}}
\usepackage{color}
\begin{document}

\title{Cosmological shock waves: clues to the formation history of haloes} 
\author[S. Planelles  \& V. Quilis] 
{Susana  Planelles $^{1,2}$\thanks{e-mail: susana.planelles@oats.inaf.it} \&  
Vicent  Quilis$^{3}$\thanks{e-mail: vicent.quilis@uv.es} \\
$^1$ Astronomy Unit, Department of Physics, University of Trieste, via Tiepolo 11, I-34131 Trieste, Italy\\
$^2$ INAF, Osservatorio Astronomico di Trieste, via Tiepolo 11, I-34131 Trieste, Italy\\ 
$^3$Departament   d'Astronomia    i   Astrof\'{\i}sica,   Universitat   de
Val\`encia,   46100   -    Burjassot   (Valencia),   Spain}  

\date{Received date / Accepted date}

\maketitle

\begin{abstract}

Shock waves developed during the  formation and evolution of cosmic structures
are key features encoding crucial information on the hierarchical formation of
the  Universe.   We present  the  analysis of  an  Eulerian adaptive  mesh
refinement  (AMR)  hydrodynamical  and  N-body simulation  in  a  $\Lambda$CDM
cosmology especially focused on the  study of cosmological shock waves.  The combination
of a shock-capturing algorithm together with the use of a halo finder allows us to study
the morphological structures of the shock patterns, the statistical properties
of shocked cells,  and the correlations between the  cosmological shock waves
appearing  at different scales  and the  properties of  the haloes  harbouring
them.   According  to  their  localisation  with  respect  to  the population of haloes  in  the
simulation,  shocks can  be split into  two broad classes:  internal weak
shocks related with evolutionary events within haloes, and external strong shocks
associated with large-scale events.  The shocks segregation  according with their
characteristic sizes  is also visible  in the shock distribution  function. This
function contains information on the  abundances and strength of the different
shocks, and it can be fitted by a double power law with a break in the slope around
a Mach  number  of  20.   We  introduce  a  generalised  scaling  relation  that
correlates the  average Mach  numbers within the  virial radius of  haloes and
their  virial masses.  In  this plane,  Mach  number -  virial  mass, two  well-differentiated  
regimes appear.  Haloes  occupy different  areas  of such
plane  according  to their early  evolutionary  histories:  those haloes  with  a relatively  quiet
evolution have an  almost constant Mach number independently  of their masses,
whereas haloes undergoing significant merger events very early in their 
evolution show a linear dependence with their masses.  At
high redshift, the  distribution of haloes in  this plane forms an  L-like pattern   
that evolves with time  bending the vertical  branch towards the
horizontal one.  We  prove that this behaviour is  produced by haloes reaching
low average  Mach numbers  as they evolve  towards a virialised  state.  The
analysis of  the propagation speed  and size of  the shock waves developed around  haloes
could give some hints on the formation  time and  main features of the
haloes.  The  possible future detection  of all sort  of cosmic shocks  by the
forthcoming telescopes, such as the  SKA, could open a  new window to indirectly
unravel some of the main features of haloes and, therefore, the formation process of 
the cosmic structures.

\end{abstract}
 

\section{Introduction}

The  cosmological shock  waves develop  as a  consequence of  the hierarchical
formation of structures in the Universe and, therefore, they are crucial ingredients in
a unified  picture of  the formation of  cosmological structures.  In  a first
phase, gravitational  energy associated to  the collapse of dark  matter haloes
corresponding  to galaxy clusters  and galaxies  is transformed  into internal
energy  of the  intra cluster  medium (ICM)  and inter  galactic  medium (IGM)
gaseous components. In the following phases, the evolution of those structures
also  produces mergers  and  accretion phenomena  that  modify the  energetic
balance of  the gas  through shock waves.  Therefore, the shocks  associated to
cosmic structure  formation and  evolution encode information  about the 
formation of the structures  and their thermal impact on the gas.

The cosmological shocks  can be broadly classified into  two categories \citep[e.g.,][]{ryu03}:
external and internal.  External shocks surround filaments, sheets,
and haloes, while internal shocks  are located within the regions bound by
external  shocks and  are created  by flow  motions correlated  with structure
formation and  evolution.   On large  scales,  the thermal  history of  galaxy
clusters is  dominated by the infall  of material onto dark  matter haloes and
the conversion  of gravitational energy into thermal energy  of the
gas.  This process occurs through the heating of the gas via strong (external)
accretion    shocks     surrounding    galaxy    clusters     and    filaments
\citep[e.g.,][]{ryu03,  miniati01, pfrommer06}.  Inside  collapsed structures,
weaker (internal)  shocks can  be subdivided in  three different  classes: (i)
accretion shocks caused by  infalling gas in cosmic structures,
(ii) merger shocks resulting from merging haloes, and (iii) random flow shocks
inside  nonlinear structures produced  during hierarchical  clustering.  These
internal shocks contribute to the virialization of haloes.

The role of  shocks in cosmological structures has  been studied from different
complementary  approaches. From the  observational point  of view,  strong
shocks usually develop in the  external low-density regions of galaxy clusters
where observable emission, such as  X-ray emission, is weak.  As  a consequence, from  
an observational
point of  view, detecting shocks on large scales is still very  challenging.  
Despite these difficulties, few large cosmic shocks have been positively detected
by means of the so-called radio relics 
\citep[e.g.,][]{ensslin98}, commonly defined as elongated radio emission not associated with the 
cluster centre or an active cluster radio galaxy. These radio relics are found in $\sim 30$
clusters \citep[e.g.,][]{bagc06, venturi07, bona09, vanw09}. 
Peripheral  radio relics on the outskirts of some massive clusters are the brightest 
of the cluster shock structures and are interpreted as  externally propagating merger shocks
that reach higher Mach numbers\footnote{The
  Mach  number, which  characterises the  strength  of shocks,  is defined  in
  Eq.~\ref{eq:defmachnum}.} as they enter the low-density regions of the ICM.  
Concerning internal shocks, 
in a few cases,   shocks driven by merging events
have been  observed with very low ($\approx  1.5-3$) Mach numbers
\citep[e.g.,][]{marke99, marke02, marke07}.

From the theoretical point of view,  several attempts to study shocks were done
using   semi-analytical   approaches   \citep{press74,   ST,   pav06,fujita01,
  gabici03}. However, the complexity of the  considered scenario proved the use of
numerical approaches as the most  suited to study and characterise the shocks
produced during  the formation and  evolution of  cosmic  structures. There
have  been numerical  studies of  shocks using  both  Eulerian, ``single-grid"
\citep[e.g.,][]{quilis98,    miniati_etal00,ryu03, kang07, vazza09}    and
``AMR-grid"   approaches   \citep{skillman08, vazza09b, vazza10},    
as   well   as   SPH   codes
\citep[e.g.,][]{pfrommer06,pfrommer08, hoeft08}.  Even   when  the  debate   about  the
benefits and drawbacks  of  different numerical techniques 
is  still open \citep[e.g.,][]{agertz07}, the inherent shock-capturing  
properties of the Eulerian techniques based on Riemann solvers, 
able to deal with shocks  by construction, seem to position this
kind of tools as the recommended ones when tackling shock related processes.
In this regard, \citet{vazza11} presented a numerical study of non-radiative cosmological simulations, at
various resolutions and performed with different cosmological codes, comparing 
the properties of thermal gas and shock waves in large scale structures.
Even when the bulk of thermal and shock properties  were reasonably in agreement 
between the Eulerian and the SPH codes, they also reported some significant 
differences between them like, for instance, differences of large factors ($\sim 10-100$) 
in the values of average Mach numbers and shock thermal energy flux in the most 
rarefied regions of the simulations, significantly different 
phase diagrams of shocked cells in grid codes compared to SPH, or 
sizable differences in the morphologies of accretion shocks between grid and SPH methods.

Pioneering attempts  to characterise  shock waves in  cosmological simulations
were carried out by  \cite{quilis98} and \cite{miniati_etal00}.  They employed
fix grid  Eulerian simulations and shock-detecting schemes based  on jumps in
the  main  thermodynamical  quantities.   Later  works  adopted  more  refined
shock-detecting  algorithms and  were more  focused onto  the  distribution of
energy  dissipated  at  shocks  \citep[e.g.,][]{miniati02, ryu03,  pfrommer06,  vazza09}.
Despite the important advances, in  the first works using uniform grid-based
codes,  it was not  possible to  cover the  spatial resolutions  required to
describe both the complex flows within haloes and their coupling to large-scale
structures.  Further improvement has been reached by \cite{skillman08} using a
shock-detecting scheme looking for  shocks in the direction of the temperature
gradients on an AMR grid.

These numerical simulations have begun to reveal  a rich network of shock structures
throughout the ICM 
\citep[e.g.,][]{miniati01b, ryu03, pfrommer08, batt09, skillman10, vazza10}.
However, in spite  of all  previous works, the  identification and  characterisation of
shocks  is still  challenging  due to  the  complex dynamics  involved in  the
formation and  evolution of  cosmological structures and to the large dynamical 
range needed to describe all the scales involved by shocks.    

Our purpose  in the present
paper is to pursue the analyses of the main properties of the shock waves developed
during the evolution of a high resolution hydrodynamical and N-body simulation
of a  large cosmological volume performed  with an AMR  cosmological code.  
In addition, we will put especial emphasis in analysing the existing connection 
between the cosmological shock waves and the population of haloes.  
In order to do so, we have  developed a numerical algorithm able of detecting and
characterising shocks in 3-D AMR simulations.  The use of AMR hydro codes turns
out  to be crucial  so as  to obtain  good dynamical  ranges with  an advanced
hydrodynamical algorithm that is able to capture shocks very accurately.

The paper is organised as follows. In Section 2, we present the
technical details  describing the simulation and the shock-finding algorithm.
The results of the analysis of the simulation are presented in Section 3.
Finally,  in Section 4, we summarise and discuss our results.

\section{Detecting shock waves}

\subsection{Simulation details}

The  simulation  used  in  this paper  was  performed  with  the
cosmological  code  MASCLET \citep{quilis04}.   This  code couples  an
Eulerian  approach  based  on the so-called 
 {\it high-resolution  shock-capturing}
techniques \citep[e.g.,][]{leveque92} for describing  the  gaseous component,  with a  multigrid
particle mesh  N-body scheme for evolving  the collisionless components
(dark matter and stars).  Gas, dark matter, and stars are coupled by the gravity solver.
Both  schemes  benefit of  using  an  adaptive  mesh refinement  (AMR)
strategy, which permits to gain spatial and temporal resolution.

The numerical  simulation was run  assuming a spatially  flat $\Lambda
CDM$  cosmology, with  the following  cosmological  parameters: matter
density    parameter,    $\Omega_m=0.25$;    cosmological    constant,
$\Omega_{\Lambda}=\Lambda/{3H_o^2}=0.75$;  baryon  density  parameter,
$\Omega_b=0.045$;  reduced Hubble  constant, $h=H_o/100  km\, s^{-1}\,
Mpc^{-1}=0.73$;  power  spectrum index,  $n_s=1$;  and power  spectrum
normalisation, $\sigma_8=0.8$.

The initial  conditions were  set up at  $z=50$, using a  CDM transfer
function from \citet{EiHu98},  for a cube of comoving  side length $64
\,  Mpc$ discretised  in  $512^3$ cubical cells.

A first level  of refinement (level $l=1$) for the  AMR scheme was set
up from the initial conditions by selecting regions satisfying certain
refining criteria,  when evolved  -- until non-linear phase--  using the
Zel'dovich  approximation.  The  dark matter  component in  the initial
refined  region was  sampled with  dark matter  particles  eight times
lighter than  those used  in regions covered  only by the  coarse grid
(level $l=0$).   During the evolution, regions on  the different grids
are refined  based on  the local baryonic  and dark  matter densities.
Thus, a cell is refined, independently of the refinement level, if its 
dark matter (gaseous) mass is larger than 
$4.74\times10^8M_\odot \, (1.04\times10^8M_\odot )$. This is equivalent 
to refine the cell if its density increases a factor of eight. 
The ratio  between the cell  sizes for a  given level ($l+1$)  and its
parent   level  ($l$)   is,   in  our   AMR  implementation,   $\Delta
x_{l+1}/\Delta  x_{l}=1/2$.  This  is a  compromise value  between the
gain in resolution and possible numerical instabilities.

The simulation presented  in this paper uses a  maximum of six levels
($l=6$) of refinement, which gives  a peak physical spatial 
resolution of $\sim4\,  kpc$. For the 
dark  matter we  consider two  particles species
which correspond to the particles on the coarse grid and the particles
within the first  level of refinement at the  initial conditions.  The
best  mass resolution  is  $\sim 2.7\times  10^8\, h^{-1}\,  M_\odot$,
equivalent to distribute $512^3$ particles in the whole box.

Our simulation includes cooling  and heating processes which take into
account inverse Compton and free-free  cooling, UV 
heating \citep{hama96} at $z \sim 6$, atomic and molecular cooling for a 
primordial gas, and star formation. In order to compute
the abundances  of each species ($H, \, He, \, H^+, \, He^+,\, He^{++}$), 
we  assume that the  gas is optically
thin  and in ionization  equilibrium, but  not in  thermal equilibrium
\citep{katz96,theuns98}.  The tabulated  cooling rates were taken from
\citet{sudo93} assuming a constant  metallicity 0.3 relative to solar.
The cooling curve was  truncated below temperatures of $10^4\,K$.  The
cooling and heating  were included in the energy  equation 
\citep[see Eq.~3 in][]{quilis04} as extra source terms.

The star  formation is  introduced in the  MASCLET code  following the
ideas  of \citet{yepes97}  and \citet{springe03}.   
In  our particular
implementation, we assume that cold  gas in a cell is transformed into
star  particles on  a  characteristic time  scale  $t_*$ according  to
$\dot{\rho_*}=-\dot{\rho}=(1-\beta)\,{\rho}/{t_*(\rho)}$  where $\rho$
and  $\rho_*$  are the  gas  and  star  densities, respectively.   The
parameter  $\beta$  stands for  the  mass  fraction  of massive  stars
($>8\,M_{\odot}$) that explode as  supernovae, and therefore return to
the  gas  component in  the  cells.   We  adopt $\beta=0.1$,  a  value
compatible with a Salpeter IMF.  For the characteristic star formation
time,        we        make        the        common        assumption
$t_*(\rho)=t^*_o(\rho/\rho_{_{th}})^{-1/2}$,       equivalent       to
$\dot\rho_*=\rho^{1.5}/t^*_o$   \citep{keni98}.   In   this   way,  we
introduce a dependence on the local  dynamical time of the gas and two
parameters, the density threshold for star formation ($\rho_{_{th}}$)
and  the corresponding  characteristic time  scale ($t^*_0$).   In our
simulation,       we        take       $t^*_0=2\,       Gyr$       and
$\rho_{_{th}}=2\times10^{-25}\,g\,cm^{-3}$.  From  the energetic point
of view,  we consider that each  supernova dumps in  the original cell
$10^{51}\, erg$ of thermal energy. 

In the practical implementation,  we assume that star formation occurs
once every global  time step, $\Delta t_{l=0}$, and  only in the cells
at the highest level of refinement.  Those cells at this level
of   refinement,  where  the   gas  temperature   drops  below   $T  <
2\times10^4\,    K$, the divergence of the velocity of the
 gas is $\nabla \cdot  {\bf v} < 0$,    
and   the    gas    density    is   $\rho    >
\rho_{_{th}}=2\times10^{-25}\,  g\,  cm^{-3}$,  are suitable  to  form
stars.   In  these  cells,  collisionless  star  particles  with  mass
$m_*=\dot\rho_*\Delta  t_{l=0}\Delta x_l^3$ are  formed.  In  order to
avoid sudden changes in the  gas density, an extra condition restricts
the    mass    of    the    star    particles    to    be    $m_*={\rm
min}(m_*,\frac{2}{3}m_{gas})$, where  $m_{gas}$ is the  total gas mass
in the considered cell. 

According with the previously described scheme for the star formation in 
our simulations, we include only one simple case of feedback, that is, the 
 contribution in mass and energy from type II supernovae. No feedback 
from active galactic nucleus (AGN) or 
type Ia supernovae are considered. Although this point could be 
improved, how to model AGN feedback in cosmological simulations is 
still a matter of discussion \citep[e.g.,][]{Fabjan2010}. 
Besides, we do not expect a dramatic change
in our results linked with the inclusion or not of feedback, since the formation and 
evolution of the cosmological shock waves are supposed to be
mainly driven by the assembly of cosmic structures and, therefore, 
gravity should be the main ingredient to model.

A final technical consideration concerning the numerical resolution of the simulation must be done. We have 
not performed any resolution test and assume that our numerical resolution is adequate to study the processes related with shocks. We base our assumption in the work by \cite{vazza11}, where authors compare the properties of shocks using different cosmological codes. We must point out that our numerical resolution -- in cell size and particles masses -- is much better than in all the simulations considered in Vazza et al. (2011) (see Table 1 of this reference), where the authors conclude    that
 a very good convergence, better than a $\sim 10$ per cent
level,  in the most important shock statistics is expected for spatial resolutions of $\sim 50-100$ kpc, which are above than our best resolution.

\subsection{Basic relations}

Shocks  produce irreversible  changes in  the gas of the cosmic structures.   
As a
consequence, the formation of a shock in a cosmological volume 
produces a jump in
all the thermodynamical  quantities. If we assume that  the pre-shocked medium
is  at  rest  and in  thermal  and  pressure  equilibrium, the  pre-shock  and
post-shock values  for any of  the hydrodynamical variables  are unambiguously
related  to the  shock Mach  number,  $\mathcal{M}$.  The  Mach number,  which
characterises the strength of a shock, is given by
\begin{equation}
\mathcal{M}=v_{s}/c_{s},  
\label{eq:defmachnum}
\end{equation}
where $v_{s}$ is the shock speed  and $c_{s}$ is the sound speed
ahead of the shock.

All  the information  needed to  evaluate  $\mathcal{M}$ is  contained in  the
Rankine-Hugoniot jump conditions.  If the adiabatic index is  set to $\gamma =
5/3$ we obtain, for the density ($\rho$), the temperature (T), and the entropy
($S=T/\rho^{\gamma-1}$), the well-known relations \citep[e.g.,][]{landau66}:

\begin{equation}
\frac{\rho_{2}}{\rho_{1}}=\frac{4\mathcal{M}^{2}}{\mathcal{M}^{2}+3}  
\label{eq:dens}
\end{equation}

\begin{equation}
\frac{T_{2}}{T_{1}}=\frac{(5\mathcal{M}^{2}-1)(\mathcal{M}^{2}+3)}{16\mathcal{M}^2}
\label{eq:temp}
\end{equation}

\begin{equation}
\frac{S_{2}}{S_{1}}=\frac{(5\mathcal{M}^{2}-1)(\mathcal{M}^{2}+3)}{16\mathcal{M}^{2}}
(\frac{\mathcal{M}^{2}+3}{4\mathcal{M}^{2}})^{2/3}
\label{eq:entropy}
\end{equation}
with indices $1,2$ referring to pre- and post-shock quantities, respectively.  

The Mach  number can be obtained from  the jumps in any  of the hydrodynamical
variables  (Eqs.~\ref{eq:dens}--\ref{eq:entropy})  or  from a  combination  of
them.  In  the case of relatively large  Mach numbers, since the  value of the
density jump saturates at $\rho_2/\rho_1=4$ (Eq.~\ref{eq:dens}), strong shocks
cannot be detected  from density jumps.  As a  consequence, the most effective
methods to measure $\mathcal{M}$ are those considering temperature and entropy
jumps.

\subsection{Shock-finding algorithm}

Any shock-finding method relies on accurately identifying and
quantifying  the  strength of  shocks.   Although, during  the  
simulations using an 
Eulerian approach, the shocks  are
automatically  detected  by  the  Riemann  solver  within  the  hydrodynamical
routine,  the analysis  of  these  shocks requires  additional
considerations.
 
Recent works on algorithms to find and track shocks in grid simulations have stablished two 
procedures. Both methods require a previous step to mark the cells which are involved in a shock 
at the time corresponding to the analysed output. Once the shocked cells have been identified, 
the two methods differ in how to define the shocks. The first method  \citep[e.g.,][]{vazza09}
relies on  following temperature or velocity jumps across the previously tagged 
shocked cells along the different  coordinate axes. 
We define this method as the coordiante or directional-splitting method. 
According with \citet{vazza09}, the results seem to be quite insensitive to the use of temperature or velocity jumps, 
with minor differences in the most rarefied environments.
This kind of methods has a minor drawback when examining shocks  
that do not propagate along the coordinate axes.
The second group of methods to find shocks use the local temperature gradient to figure out 
the direction of the shock propagation \citep[e.g.,][]{skillman08}. These methods would be more precise 
when dealing with  complex flows or weak shocks.
So as to minimise a possible smearing of the shocks produced by the coordinate-splitting or 
other uncontrolled numerical effects, the so-called coordinate-splitting methods define the final Mach number as the average 
of the Mach numbers along each coordinate axis.  With this implementation, the differences between schemes based on a precise estimate of the direction of the shock propagation and others based on the directional-splitting are not expected to be significative.

Our shock-finding algorithm is based on temperature jumps and, therefore, comparable to the method in \citet{vazza09}.  
Thus, the main steps of the shock-finding algorithm used in this paper are the following:

\begin{enumerate}

\item  All cells  within  the
  computational volume are classified as tentative shocked or not shocked.  
  A cell is labelled as tentative
  shocked if the fluid inside the cell meets the following requirements:
\begin{eqnarray}
\nabla \cdot {\bf v} < 0 \\
\nabla T \cdot \nabla S > 0 
\end{eqnarray}  
where  ${\bf v}$,  $T$  and $S$  are,  respectively, the  velocity field,  the
temperature and the entropy of the fluid (gas) within the cell.

\item Among all the tentative shocked cells,  
  the cell where $\nabla \cdot {\bf v}$ is minimum 
  is flagged as the first shocked cell. 
  
 \item The sense of shock propagation with respect to the computational grid  is computed by 
 moving outwards from
 the previously identified shocked  cell along  each of the  three coordinate axes. 
 This extension  is done meanwhile 
 there are tentative shocked cells and the temperature and  the density in
  these tentative shocked cells satisfy that $T_2 > T_1 $\ and $\rho_2 > \rho_{1}$.  
  The subscripts 1 and 2 refer to the tentative shocked cells ahead and behind of the central shock 
  cell in the shock reference frame, respectively.
 The shock discontinuities caught in the simulations by this method
are typically spread over a few cells.

\item  Once  the  furthest  shocked cell ahead 
and behind of the central cell of the shock -- that we denote as  the pre-  and post-shock  cells --  in  
each  coordinated
  direction  are   found,  the  temperatures  $T_1$  (pre-shock)   and  $T_2$
  (post-shock)  of these cells are taken  and  three  different Mach  
  numbers  (one for  each direction) are calculated from Eq.~\ref{eq:temp}.

\item The strength of the shock is computed  by combining the Mach numbers  
measured along the three coordinate axes:
  $\mathcal{M}=(\mathcal{M}_{x}^{2}+\mathcal{M}_{y}^{2}+\mathcal{M}_{z}^{2})^{1/2}$,
  which   minimises   projection   effects   in  case   of   diagonal   shocks
  \citep[e.g.,][]{ vazza09}.
  
\item The shocked cell drops out of the list of the tentative shocked cells, and the process 
is iteratively repeated focusing on the cell with    minimum $\nabla \cdot {\bf v}$ among the 
remaining tentative shocked cells.
 
\end{enumerate}

The whole procedure identifies  
all  the  shocked  cells  within  the  computational  box  and obtains 
their Mach number. The assembly of all  
these shocked cells defines the
characteristic shock surfaces associated to shock waves.

When this procedure is applied to an AMR simulation, 
the analysis is carried out in a hierarchical fashion, first 
on the  most highly refined grids, moving  down
to progressively coarser  levels of resolution.  Given that  this procedure is
applied, independently,  at each  level of refinement  of the  simulation, the
algorithm is able to find, in  a natural way, shock waves related to different
spatial scales provided by the simulation itself.  
Although this is a very useful and interesting feature of our approach, 
the use of an AMR grid makes the process  more complicated 
compared with a fix grid approach, especially when shocks extend 
over  several subgrids at different levels.  
In this case, the algorithm goes on looking for the boundaries of the shock wave 
on the parent subgrid -- in a lower level of refinement -- that contains the original subgrid
on which the shock was originally identified. This mechanism can be applied recursively on several levels if needed. 
 
In all the analysis  performed in following sections of this paper, in order 
to avoid  noisy shock patterns with very  low Mach numbers,  we have  
considered a  Mach number  minimum threshold equal to $1.3$.

\begin{figure*}
\centering\includegraphics[width=16 cm]{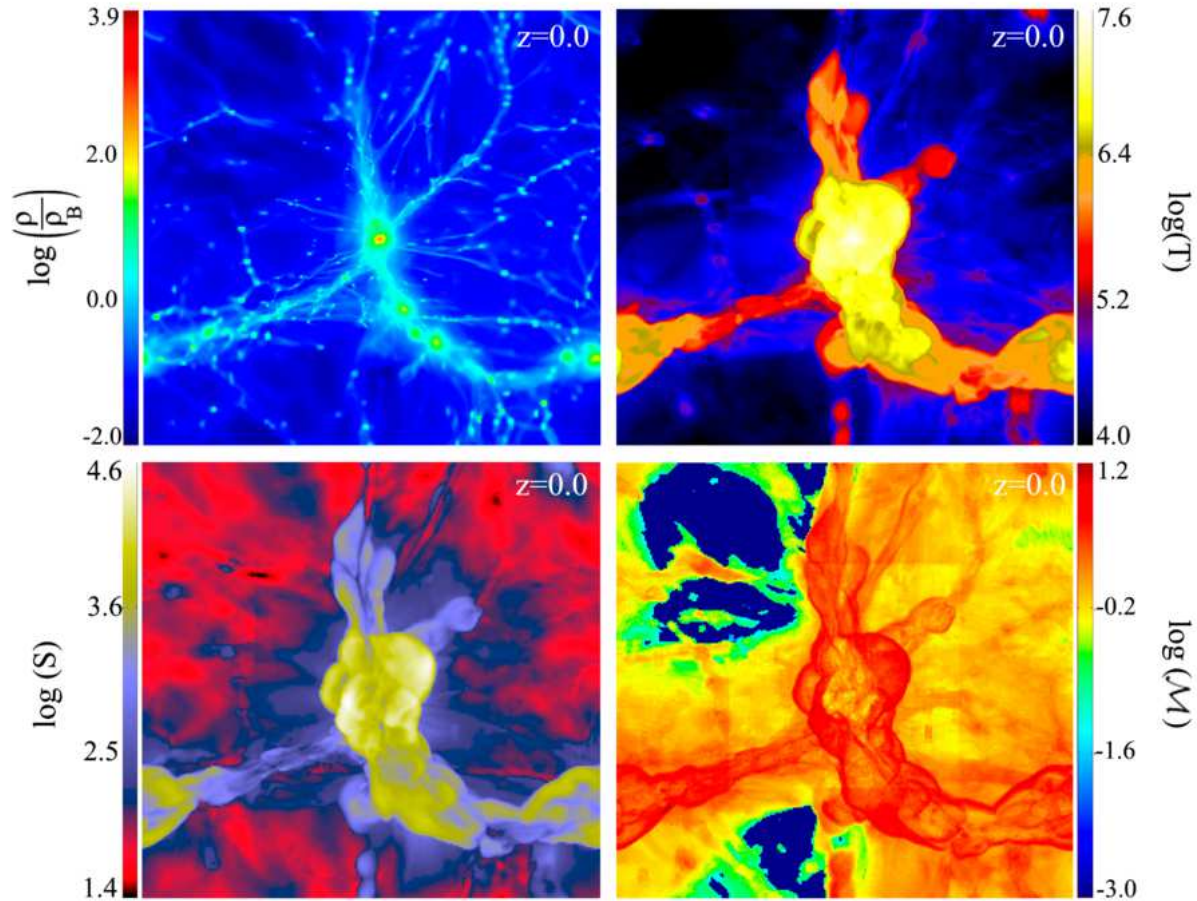}
\caption
 {Large-scale distribution of several quantities at $z=0$. Each two-dimensional  panel is  
 $64$ Mpc side length. All  quantities are
 integrated along a slice of thickness  $L=10$ Mpc according with $q_i=\frac{1}{L}\int{q dx}$ where 
 $q$ is a generic quantity to be projected and $q_i$ is the final plotted variable. The panels show:
 the integrated gas overdensity, $\rho / \rho_{_{B}}$, being $\rho_{_{B}}$ the background density  (upper left), the integrated gas temperature in K
 (upper right), the integrated gas entropy in $keV\, cm^2$ (lower left) and the integrated Mach number 
 (lower right). Different palettes have 
been used in order to highlight the particular features of each panel.}
\label{fig:panel1}
\end{figure*}

\section{Results}

\subsection{Distribution of large-scale shocks}

Shocks   fill    the   simulated   volume   in   a    very   complex   pattern
\citep[e.g.,][]{miniati_etal00,  ryu03}.   Figure~\ref{fig:panel1} illustrates
typical  structures  found  in  large-scale  cosmological  simulations.   The
different two-dimensional panels  ($64$  Mpc side
length) represent the following integrated quantities along slices  of $10$  Mpc thickness at $z=0$: 
the  gas overdensity (upper  left), the  gas temperature
(upper right), the gas entropy (lower  left) and the Mach number (lower right).  
All the panels  show the logarithm  of the different
quantities.  
The four projections are  centred at the position of the largest
cluster in  the simulation\footnote{See Section 3.2 for further details on the halo population
obtained in the present simulation.}  
($M_{vir} \sim 8.0  \times10^{14}\, M_\odot$ and $R_{vir} \sim 2.4$ Mpc)
which is  almost at  the centre  of the  box.  

The panel presenting the gas overdensity in  Fig.~\ref{fig:panel1}
shows the common picture of galaxy clusters connected through filaments.
The shock waves, if existing, are not appreciable in this picture. The panel 
displaying the gas temperature clearly shows high temperature regions, where the
gas has been heated up by shock moving outwards from the centre of the 
structure, and low temperature spots associated to regions where the gas 
cooled down during the cosmic evolution. The temperature plot gives a first preliminary
idea of the regions which have been shocked. The entropy panel, as it would be 
expected, is a step forward highlighting the complex structure of shocks. In this 
particular panel, the boundaries of high temperature regions appear as high entropy
zones indicating the presence of a shock or strong gradient. For completeness, the 
last panel shows the Mach number associated to the studied region. Now, 
the complex
structure of shocks is clearly visible, showing a remarkable correlation with the 
entropy panel, although much sharper in this case. The Mach number map also allows 
to distinguish between high-Mach number shocks, wrapping the larger 
structures, and the low-Mach number shocks associated with internal small 
scale structures.

\begin{figure*}
\centering\includegraphics[width=16 cm]{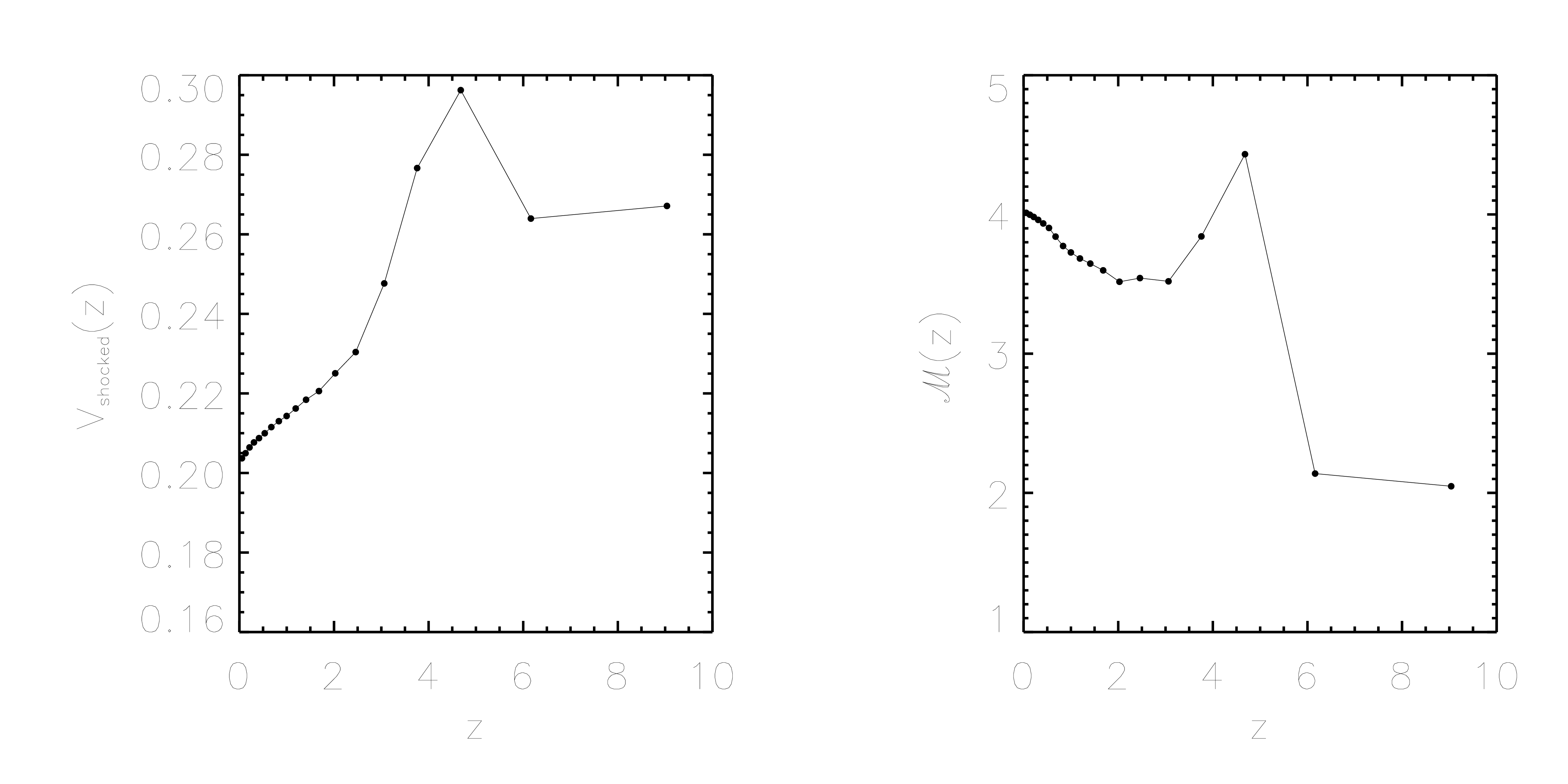}
 \caption
 {{\it Left}: Volume {\bf fraction} occupied by shocked cells in the simulation as a function of
 redshift. {\it Right}: mean volume-weighted Mach number as a function of redshift.}
\label{fig:V_M}
\end{figure*}

The cosmic  evolution of shock  strengths encodes a valuable  information about
the  thermal  history of  the  baryonic  component  of the  Universe.  In 
 Fig.~\ref{fig:V_M}, the filling volume factor of shocks -- that is,  
 the fraction of the computational volume occupied by shocked cells -- 
 is plotted versus redshift evolution (left panel), whereas the right panel shows 
 the volume-weighted average Mach number for the same time evolution.
 A warning notice must be done at this point because, as  already mentioned in Sec. 2.1, a resolution 
 test has not been performed.
 Therefore, these results could be affected by resolution effects. 

Early evolution ($z\gtrsim6$) of the shocked volume evidences a non-negligible 
fraction of the computational box involved in shocks ($\sim 27\%$). 
This quantity 
increases towards its maximum at $z\sim5$ when a step decline
of the filling factor of shocks starts. The explanation of such behaviour
is associated with huge amounts of gas falling into the potential wells created
by cosmic structures --  at different scales -- at the beginning of their non-linear evolution. As 
the gas flows have  relatively low densities and moderate velocities, 
they can easily reach supersonic regimes. Therefore, they can occupy an 
important volume of the simulated region. However, as it is shown in the right panel,  
these are weak shocks characterised by low volume-weighted Mach numbers.

As the non-linear evolution of cosmic structures continues, the number of 
collapsing regions increases, as well as the densities and velocities of 
the gas flows falling into such structures. The combination of these factors 
produces a maximum of the shocked volume and the shock strength around 
$z\sim5$.  

After this time two different trends appear depending on whether we look at 
the shocked volume or at the average shock strength. The first one continuously 
declines mainly due to the fact that structures start to collapse and to form 
bound haloes where the gas component is locked up and, therefore, the sizes 
of regions where shocks can appear get smaller as the time evolution 
goes on. On the other hand, the average shock strength shows the same 
behaviour immediately after its maximum but it changes its derivate
around $z\sim2$. The reason of this behaviour is strongly correlated 
with the increase of the merging rate among the already formed haloes of 
different masses and sizes. 
The final fraction of  the simulated shocked volume reaches  a value  of $\sim
20\%$ at  $z=0$.  

This result on the temporal evolution of the fraction of the computational volume occupied by shocked cells
seems to be in broad agreement with the results shown in  \cite{vazza09} who
obtained that the fraction of the simulated volume being shocked goes from 
 roughly $30\%$  at high redshift ($z\ge 6$) down to $\sim 15 \%$ at $z=0$.

\begin{figure}
\centering\includegraphics[width=8 cm]{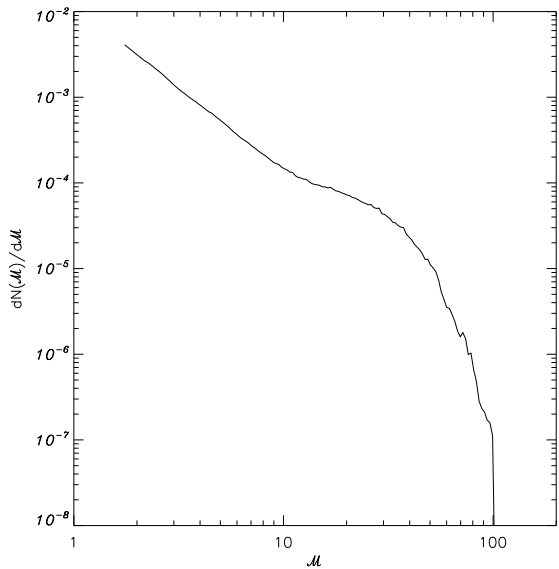}
\caption
{Shocks distribution function (SDF) at $z=0$ as a function of their Mach
 number in logarithmic bins.}
\label{fig:mass_func_shocks}
\end{figure}

A crucial tool in the study of cosmic structures is the mass distribution 
function of 
the different objects in the Universe. Following a similar approach to the
mass function, we study the   shocks distribution  function (from now on SDF), 
defined as 
the variation of the number of shocked cells (volume-weighted)
as a function of 
the Mach number. In Fig.~\ref{fig:mass_func_shocks}
the differential  SDF at $z=0$ is plotted 
-- in logarithmic scale --  to show 
the distribution of shocks in the simulated volume.

The most remarkable feature of the SDF is the presence of two clearly 
differentiated regimes with different slopes. The first regime corresponds
to low Mach numbers which are much more abundant. The second regime is
steeper than the previous one and stands for strong shocks with higher Mach 
numbers which are the most rare in the studied volume. The transition between
the two regimes takes place around $\mathcal{M}\sim 20$. This value for 
$\mathcal{M}$, as it will be discussed in the next sections, turns out to be a 
crucial number separating the internal shocks -- located within the virial 
radius of the structures -- and the external shocks at outer regions. 
The two different regions of the SDF can be fitted by power laws of the form 
$dN(\mathcal{M})/d\mathcal{M}    \propto\mathcal{M}^{\alpha}$.  
Thus, for low Mach numbers (up to $\simeq  20$) we obtain a
slope of $\alpha \simeq -1.7$,  whereas for stronger shocks a steeper relation, 
$\alpha \simeq  -4.1$, is found. 
Previous works already proved that the bulk of shocks in the universe
is dominated by relatively weak shocks $(\mathcal{M}\le 2)$,
 but there is also a population of stronger shocks located in the external 
 regions of haloes where structures are not completely virialised
 \citep[e.g.,][]{ryu03, pfrommer06, vazza09}. 
 
We have also followed the time evolution of the SDF at several redshifts, 
finding very similar trends at all the epochs. The most relevant 
difference between  the distributions obtained  at different
times is  found at the high-Mach end of  the relation,
 which slowly  moves down to
lower values when  the redshift decreases. 
Only when quite high redshifts
($z\ge  6$) are reached,  an  important  change  in   the  shape  of  the
distribution is appreciated although, probably, the information provided by the 
SDF at such high redshift is not relevant due to the early stage of evolution 
of the cosmic structures.

\subsection{Shock waves and cosmic structures}

\begin{figure*}
\centering\includegraphics[width=16 cm]{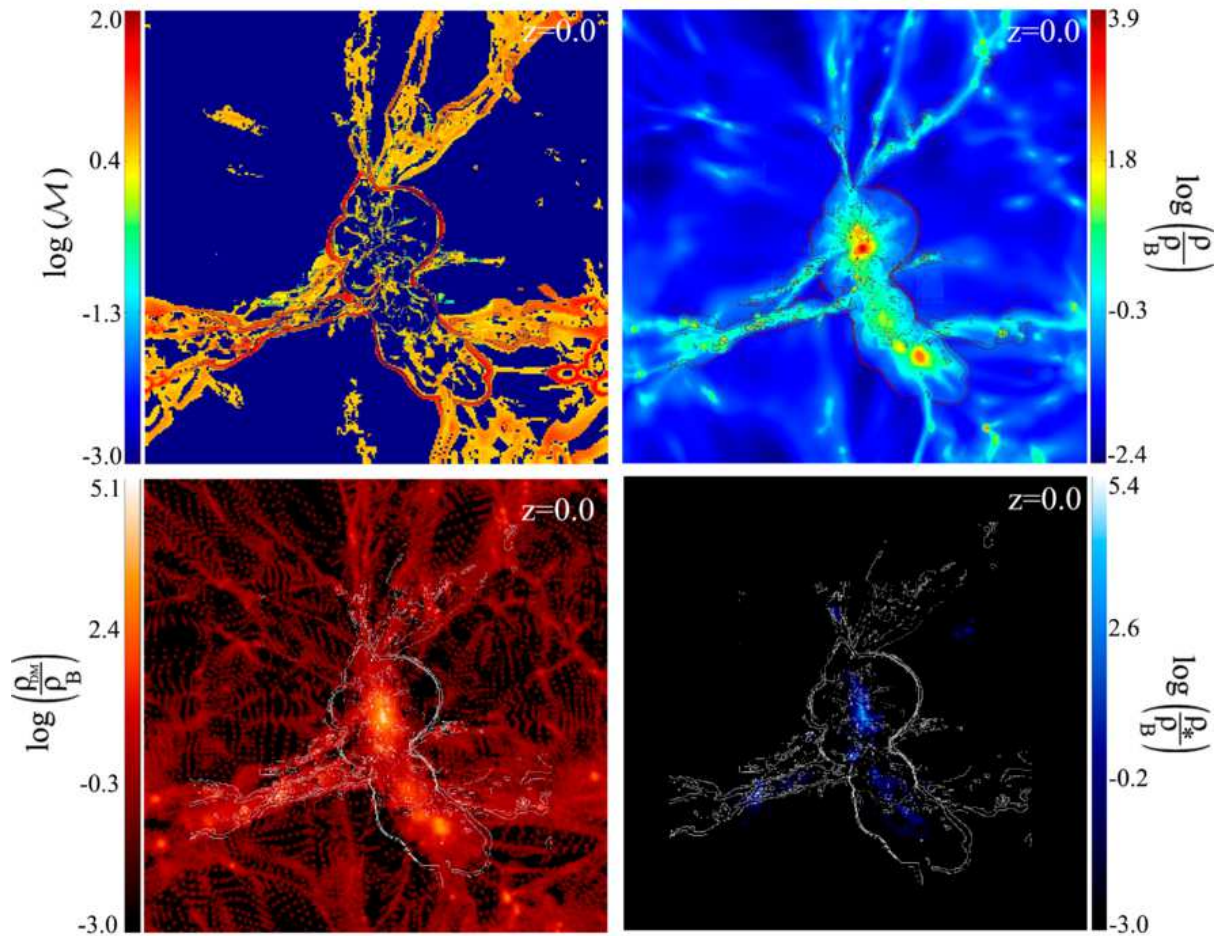}
 \caption
 {Distributions of Mach number compared 
 with dark matter $(\rho_{DM}/\rho_{_{B}})$ , gas $(\rho/\rho_{_{B}})$  and stellar $(\rho_*/\rho_{_{B}})$  overdensities at $z=0$. 
 Each panel is a thin slice of $0.2$ Mpc thickness and $64$ Mpc side length. 
 They show the Mach number distribution (upper left) and the gas, 
 dark matter and 
 stellar overdensities (upper right, lower left, and lower right panels, respectively).
In all the panels, 
the contours of the shock waves with high Mach numbers are overplotted.}
\label{fig:contornos}
\end{figure*}

Cosmological shocks derived from the formation and evolution 
of cosmic structures have to present some correlations with the main features
of the distribution of such structures. In order to deepen in this 
connection, in the present Section we discuss the correlation between the 
distribution of cosmological shock waves and the halo population
(galaxies and galaxy clusters). 
 
The first analysis one can think about is to compare the shock pattern with
the density distribution of the different components forming the haloes in 
the simulation. Thus,   
Fig.~\ref{fig:contornos} shows a 2-D projection along the z axis of the Mach
number distribution (upper left panel) to compare with 
 the gas $(\rho/\rho_{_{B}})$,  dark matter $(\rho_{DM}/\rho_{_{B}})$ and 
 stellar $(\rho_*/\rho_{_{B}})$ overdensities
(upper  right, lower  left, and  lower right  panels, respectively)  at $z=0$.
Each panel represents a thin slice of  $0.2$ Mpc thickness and $64$ Mpc side length
centred  at  the  position of the most massive halo found in the 
computational box (the same  than  in  Fig.~\ref{fig:panel1}).  All  
the plotted
quantities are in logarithmic scale.  In all these panels, the contours of the
strong shock waves -- with high Mach numbers ($\mathcal{M}>20$) -- 
are overplotted.

From Fig.~\ref{fig:contornos} we can infer, as it was expected, that the shock 
pattern perfectly traces the cosmic web \citep[e.g.,][]{miniati_etal00}. 
Nevertheless, this shock pattern has two  clearly 
separated regimes. The first of these regimes gathers the high 
$\mathcal{M}$ shocks -- overplotted as contour lines in 
Fig.~\ref{fig:contornos} --  that wrap 
filaments,  sheets, and haloes. As these shocks are located out of the virial 
radius of the structures, they are classified as external shocks. These 
shocks have  quasi-spherical  geometries and can be located at distances 
of several virial radius from the centre of the structure where they 
were created  \citep[e.g.,][]{vazza09}.
In general, such external shocks are the evolved state of the accretion 
shocks formed during the collapse of haloes. 
During their movement outwards from the halo centre,
these accretion shocks interact among them creating, therefore, a more complex pattern.

Although for the sake of clarity 
the low-Mach number shocks are not plotted in  Fig.~\ref{fig:contornos}, 
moving inwards  the  virial  radius  of haloes, more  irregular  and  weaker  shocks
($\mathcal{M}\leq5$)  are formed, in line with results from previous studies
\citep[e.g.,][]{vazza09}.    
This  second regime corresponds to the 
internal  shocks  which are mainly 
associated to random flow motions and merger events within the haloes.

\begin{figure}
\centering\includegraphics[width=8 cm]{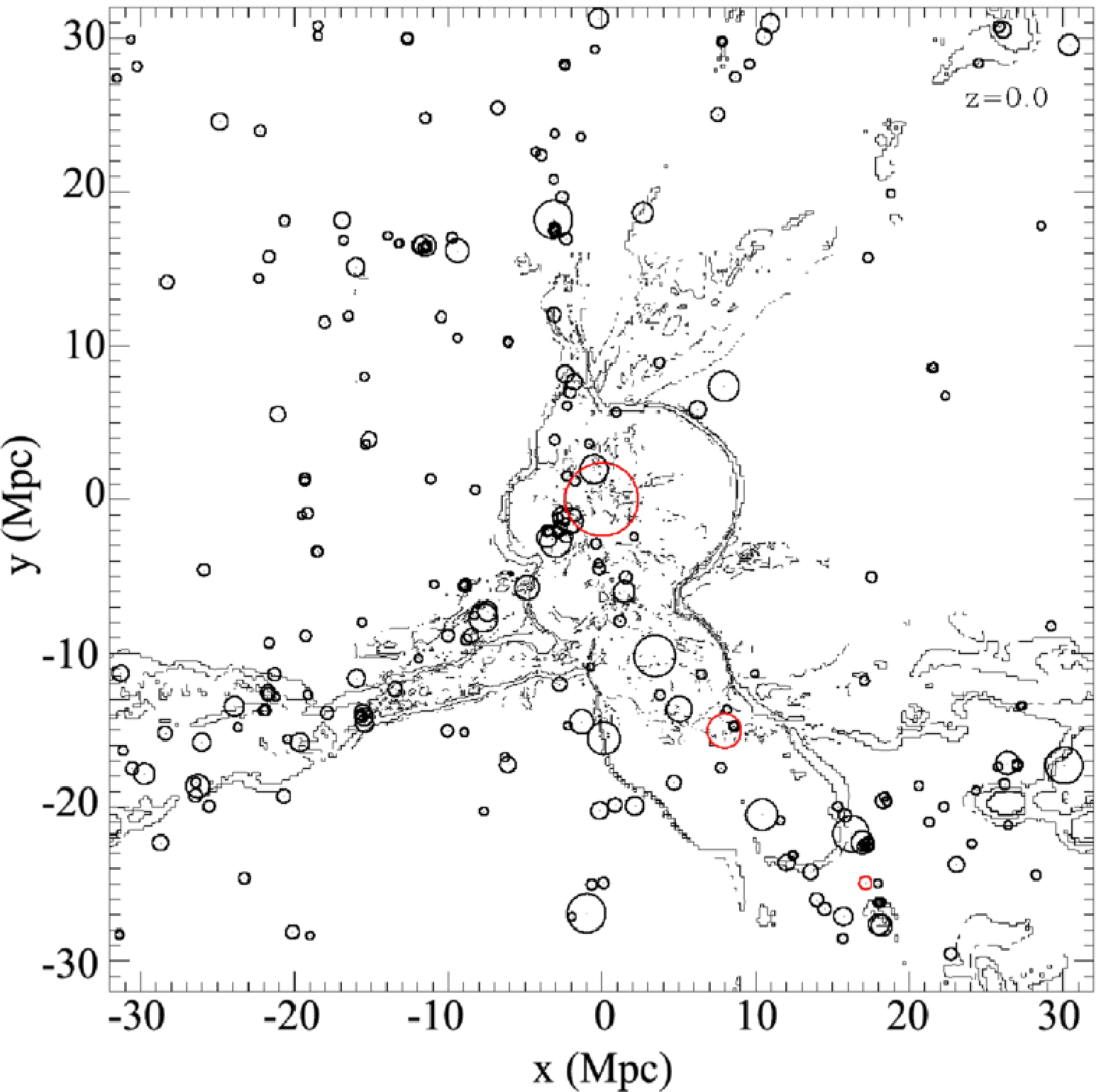}
\caption{Distribution  of dark  matter  haloes together  with  shock waves  at
  $z=0$.  The solid  lines represent six contour levels of  the 
  higher-Mach number shock waves.  This contour  has been  computed for  a slice  of the
  simulated box of $0.2$ Mpc depth and $64$ Mpc side length centred at the location of the most massive halo. 
  Circles stand for
  all the found haloes in the computational box with masses larger than $\geq10^{12}\,M_\odot$.
  We  point out  in red  those haloes  whose positions  fit in  the considered
  slice. The size of the circles represent the virial radius of the haloes.}
\label{fig:haloes_contour}
\end{figure}

In order to reinforce the previous idea,  in Fig.~\ref{fig:haloes_contour} we
compare the distribution  of shock waves with high Mach  numbers (given by the
contour lines) with  the spatial distribution of the  dark matter haloes found
in  the  simulation  at $z=0$.  This  figure  shows  the  same slice  than  in
Fig.~\ref{fig:contornos}. The  represented dark matter haloes  have been identified
with the ASOHF halo finder \citep[][]{planelles2010}. 
Only objects with virial masses\footnote{
The  virial mass  of a halo,  $M_{vir}$, is defined  as the
mass enclosed  in a spherical  region of radius $R_{vir}$  with an
average density $\Delta_c=18 \pi^2 + 82 x - 39 x^2$ times  the critical density of the Universe, 
being $x=\Omega(z)-1$  
and  $\Omega(z)=[\Omega_m (1+z)^3]/  [\Omega_m (1+z)^3 + \Omega_\Lambda]$ 
\citep{brynor98}.} 
larger than $10^{12}\,M_\odot$ are considered.
The total sample consists of $\sim$ 260 haloes at $z\sim 0$ within a range of masses between  
$1.0 \times10^{12}\,M_\odot$ and  $8.0\times10^{14}\,M_\odot$.
The dark matter  haloes are
plotted as  circles whose  sizes represent their  virial radius.  
Those haloes whose position perfectly fits 
in the analysed slice are drawn in red. 
It is clearly visible how the complex pattern of strong shocks 
 -- produced by the hierarchical evolution of the proto-haloes --
surrounds the galaxy cluster (large red circle) from which these shocks stem from.

\subsection{The Mach number scaling relation}

The scaling relations are powerful tools in the understanding of the 
processes sculpting the formation and evolution of galaxy clusters and 
groups. The common way to use the scaling relations consists in plotting, 
for a given characteristic radius, the cluster mass against the 
temperature, the entropy, or the X-ray luminosity.

\begin{figure*}
\centering\includegraphics[width=16 cm]{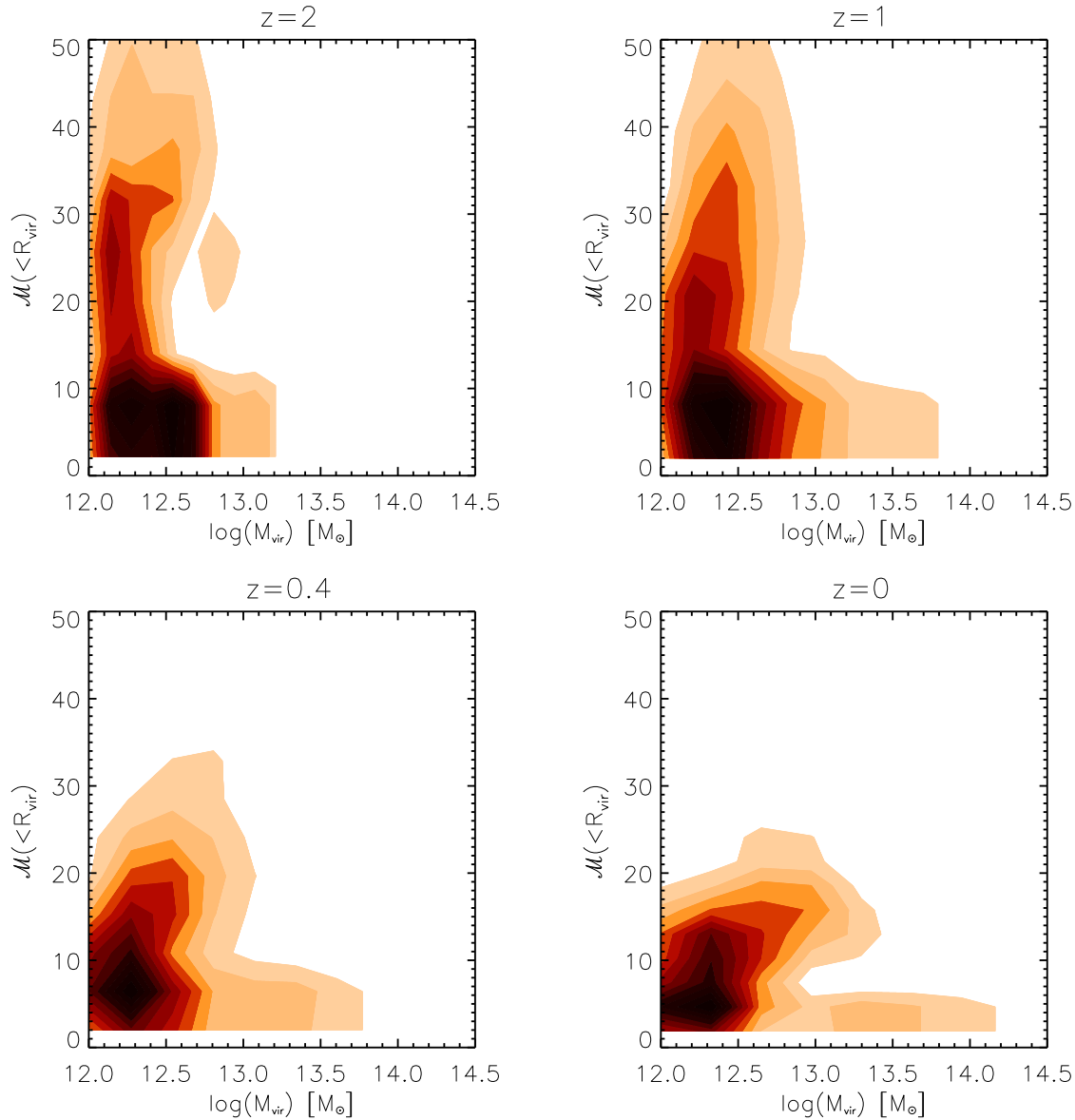}
\caption{Halo distribution in the Mach number - virial mass plane.   Results 
  at $z \simeq  2, 1, 0.4$,
  and  0,  are  shown.  The  shaded  regions  have  been computed  by  binning
  individual   haloes    into   a   two-dimensional   grid    in   the   plane
  $\mathcal{M}-M_{vir}$.  Six  contour lines  equally  spaced  are plotted  to
  highlight the two-dimensional distributions.}
\label{fig:corr_haloes}
\end{figure*}

Starting from this idea, we generalize its fundamental concept 
and we introduce  what we call the 
Mach number scaling relation. In order to build this relation we compute an average 
volume-weighted Mach number for each halo  in our sample with  
$M_{vir}>10^{12}\,M_\odot$. This volume-weighted Mach number
is computed within the virial radius of each halo and, therefore, it represents
a  virial Mach number. 
After that, we bin individual 
haloes into a two-dimensional grid of Mach number versus virial mass  
in order to have information on the number
of haloes within a certain range of Mach numbers and viral masses.
The obtained scaling relation is displayed in  Fig.~\ref{fig:corr_haloes} 
in which we show different contour lines in order 
to highlight the shape of the 2-D distribution.
Four different epochs corresponding to  $z \simeq  2, 1,  0.4$, and 0, are 
shown. The sample of haloes with  $M_{vir}>10^{12}\,M_\odot$ at these
epochs ranges from $\sim 160$ haloes at $z\simeq2$ to $\sim 260$ at $z=0$.

The analysis of Fig.~\ref{fig:corr_haloes} revels some striking features.
Let us focus first on the analysis of the distribution at 
$z=0$ (lower
right panel), where  there  are two well-differentiated trends.   
On the one hand,
there  is   an  almost   constant  region  for   low  Mach   numbers 
($\mathcal{M}\lesssim5$)  which  seems to  be  no-dependant  on  the mass  of  the
haloes. 
On the other hand,  a steeper trend along all the range of
Mach numbers seems  to be correlated with halo masses.

Our explanation for this distribution is that 
the first trend is occupied by haloes which started their evolution in 
a relatively smooth way,  with 
no mergers or a few minor mergers. The shocks within the virial 
radius producing the average Mach numbers associated with these haloes 
are related with  smooth accretion flows of gas falling into the objects 
during this quiescence evolution. Therefore,  haloes that have been quietly
set up since early phases of their evolution have 
low $\mathcal{M}$ an tend to be located at the bottom of the plane
 $\mathcal{M}-M_{vir}$ independently of their virial masses.

Complementary, haloes that initially were involved  in merger events and in active phases
of their evolution suffer violent events that produce stronger shocks 
compared with those in the previous region. In this branch of the  
$\mathcal{M}-M_{vir}$ plot, there is a strong dependence on the virial masses
of haloes.

Therefore, the evolutionary history of each halo explains 
the bimodal segregation shown in  Fig.~\ref{fig:corr_haloes}, which is also
observed through the temporal evolution. 
At early epochs, corresponding to  the formation time of large 
galaxies and small groups of galaxies ($z\sim \,2 $),
an L-like  pattern in the  plane $\mathcal{M}-M_{vir}$ appears. 
These times correspond to highly non-linear epochs of the evolution 
of the haloes, with high rates of merger events and interaction among 
the structures at different spatial and mass scales.
When advancing in time, the initial L-pattern trend progressively bends 
over to higher masses -- corresponding to big galactic haloes 
and galaxy clusters -- but lower Mach numbers, reaching the bimodal 
distribution, previously discussed, at $z\sim0$.

In order to  understand this behaviour, we have  followed the global evolution
of  individual  haloes  looking  at  their  overall  drifts  along  the  plane
$\mathcal{M}-M_{vir}$. We have studied this  evolution for the 22 more massive
haloes in the simulation which, indeed, are the best numerically resolved.
 The haloes  in this subsample  evolve according to two  different behaviours.
 Roughly the $50\%$ of this subsample  of haloes begin, at high redshift, with
 a relatively high Mach number  and evolve to progressively lower Mach numbers
 while increasing their mass. The remaining percentage of haloes depart from 
 low-Mach  number  states  ($\mathcal{M}<5$)  and tend to move,  during their evolution,
 almost parallel to the x-axis  while augmenting their masses.  Since our sample
 of haloes is  far from being statistically complete we can  not make a robust
 conclusion.  However,  our hypothesis to  explain this behaviour is  that the
 evolution  of  haloes along  the  plane  $\mathcal{M}-M_{vir}$ is  intimately
 related  with their  dynamical  history.  Like a  gross  trend, those  haloes
 suffering important  merger events (major  mergers) early in  their evolution
 only  can evolve  towards lower  Mach  numbers while  reaching an  equilibrium
 state producing, therefore,  the decline of the initial  L-like pattern into
 the flatter final  distribution. On the other hand,  haloes with a relatively 
 quiet  beginning show very low initial Mach numbers and evolve  
 without  significant changes  in  their virial  Mach
 number  and, consequently, they  move almost  parallel to  the $M_{vir}$-axis
 while increasing their mass.
 
Let us point out that this general behaviour is not in contradiction 
 with previous studies. Once haloes, due to their early evolution, 
 are initially placed in the $\mathcal{M}-M_{vir}$ plane, 
  shocks  ($\mathcal{M} \le 3$)  
 inside their virial radius can be episodically found associated to merger events
 \citep[e.g.,][]{vazza09}. 
 However, the strength of these internal shocks is not enough to break 
 the shape of the obtained L-like pattern.

Figure~\ref{fig:corr_haloes} also encloses relevant information 
on the evolution of global integrated quantities. In this regard, 
the time evolution of the mean Mach number of the
overall sample of haloes clearly shows a progressively decline
from higher  values ($\mathcal{M}\simeq10$) at  high redshifts, down  to lower
values ($\mathcal{M}\simeq5$) at $z=0$. This decrease of the mean Mach number
is fully consistent with  a picture of haloes evolving towards an equilibrium state.

\begin{figure}
\centering\includegraphics[width=8 cm]{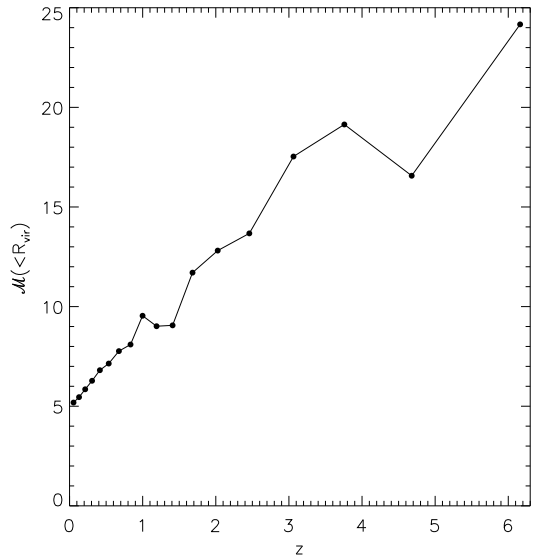}
\caption
{Mean mass-weighted virial Mach number for 
the entire population of haloes as a function
of redshift.}
\label{fig:mean_mac_z}
\end{figure}

This behaviour can be correlated with the one shown in Fig.~\ref{fig:mean_mac_z}.
In this figure, we show the evolution with redshift of the mean mass-weighted 
virial Mach number for the entire population of haloes at a given epoch. 
In this analysis we are dealing with internal shocks and, therefore, the 
Mach numbers are relatively low, going from 
$\mathcal{M} \simeq25$ at $z\simeq6$ down to $\mathcal{M}\simeq 5$ at $z=0$. 
As when discussing Fig.~\ref{fig:V_M}, it is important to mention that resolution effects could 
affect the results, especially at high redshift.

In any case,  and independently of  the considered  epoch, the  mean Mach
number of  the bulk of  haloes, as derived from Figs.~\ref{fig:corr_haloes} 
and  \ref{fig:mean_mac_z},
is always lower than  $\simeq 20-25$. 
This result perfectly
correlates with  the turn  observed in the distribution function of shocks
discussed in Sec.~3.1 (see Fig.~\ref{fig:mass_func_shocks}), 
representing the transition between  external and internal shocks.

\subsection{Halo properties and shock waves}

The results presented in Sec. 3.2 reveals the already expected conclusion that 
there should exist some sort of  
correlation between the 
external shocks surrounding clusters and galaxies and the dark matter 
haloes where such shock waves were probably originated. 
In this Section, we are going to study this correlation in detail and 
its possible implications to characterise the haloes where the shock waves are originated. 

The collapse of an isolated and idealised halo would produce an accretion shock that, after 
the formation of the halo, would move outwards from the centre of the structure defining an
spherical shell. In a more realistic scenario, and assuming that the formation of every halo produces
an accretion shock, the shocks surrounding the haloes would not be perfectly spherical and in 
some cases the formed patterns can be extremely complex \citep[e.g.,][]{miniati_etal00}. 
Nevertheless, there are situations 
where these shocks can be well defined even in a realistic case: i) for those  
haloes which have had a relatively quiet history, and 
ii) for the most massive haloes which usually
have stronger accretion shocks at their outskirts than the smaller objects. 

\begin{figure*}
\centering\includegraphics[width=16 cm]{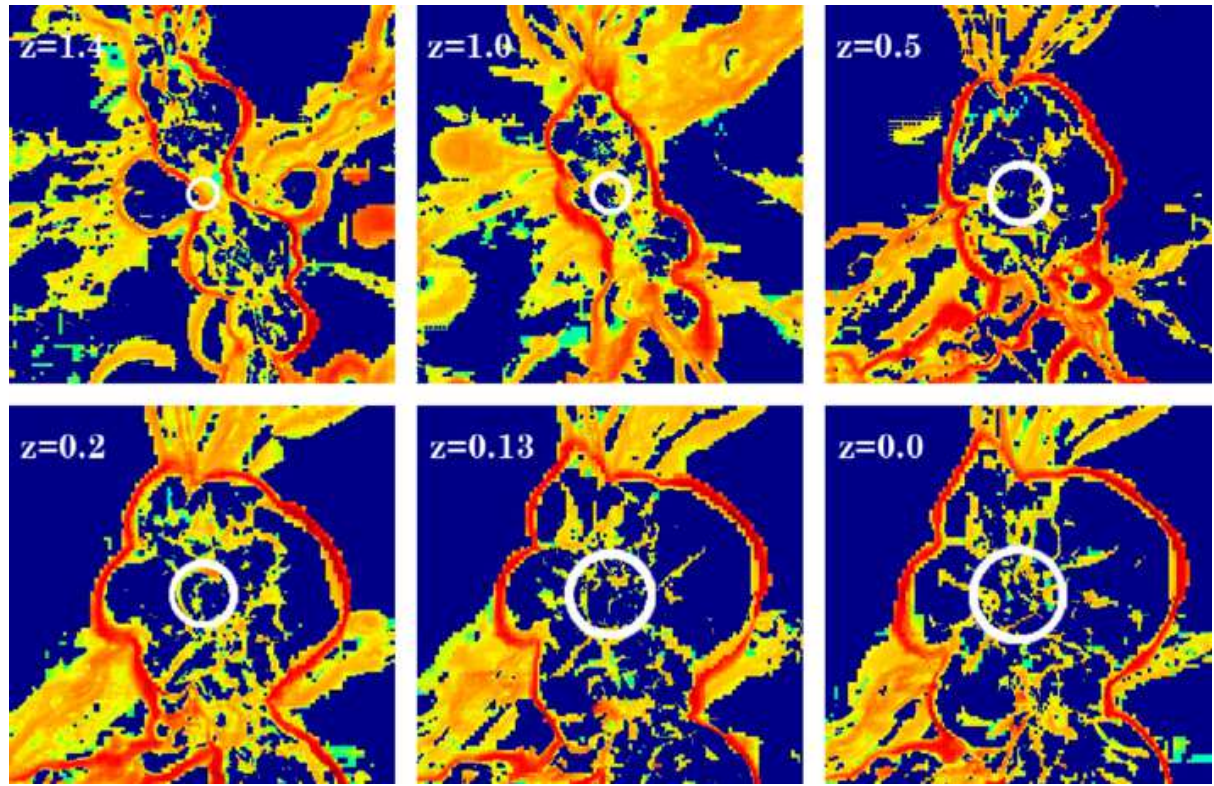}
\caption
 {Distribution of shock Mach numbers in logarithmic scale at several
 redshifts. 
 Each panel is a slice of $0.2$ Mpc thickness and $\sim$ 21 Mpc side length.
 The white circle in each panel represents the position of the main progenitor of the
 most massive halo at $z=0$. The redder the color used to plot the shock wave the higher the Mach number.}
\label{fig:secuencia_shock}
\end{figure*}

In Fig.~\ref{fig:secuencia_shock} 
we display several images 
showing the distribution of shock waves  
at several redshifts $(z=1.4, 1, 0.5, 0.2, 0.13, 0)$ around the most massive halo in the 
simulation. All the panels represent slices of $0.2$ Mpc thickness
and $\sim$ 21 Mpc side length. 
In each of the panels in Fig.~\ref{fig:secuencia_shock}, 
in addition to the shock pattern, we overplot the position and size of the dark matter halo as 
a white circle whose radius represents the halo virial radius in comoving coordinates.
 At $z=0$, the quasi-spherical accretion shock wave associated to the halo is 
 perfectly visible and easily identifiable.  By simply tracking backwards in time in the 
 different panels the position of the shock, we can arrive at the time when 
 this accretion shock was formed. If we assume that the formation of the accretion 
 shock can be defined as the formation time of the halo, this procedure produces 
 a new and natural way to estimate the formation time of a galaxy cluster, in contrast to 
 the common definition  of the cluster formation time 
\citep{lacey1993}, i.e., when the halo reaches half of its mass at $z=0$.

Besides the qualitative procedure just described, we have tried to 
track the movement of the accretion shock around the haloes in 
greater detail. In order to 
do so, we have visually identified and  measured the most
relevant quantities of the shocks. Unfortunately, the statistical limitation of 
our sample of massive haloes, plus the difficulty of the semi-manual 
methodology to follow the evolution of the accretion shocks, has 
made extremely difficult to perform this analysis and, as a consequence,  
we have been able to do it only for the most massive halo in the simulation.
In Table~\ref{tab1}, we summarise 
 the main data corresponding to the studied halo and its associated 
 external shock wave according to  Fig.~\ref{fig:secuencia_shock}.
 The measured quantities are:   
the redshift (z), the cluster virial mass ($M_{vir}$), the cluster virial 
radius ($R_{vir}$), 
the comoving radius of the shell defined by the shock ($R_{s}$), 
and the shock speed ($v_{s}$).
The radial distance from the halo centre to the shock, $R_{s}$, is estimated using a visual mechanism which can 
only produced approximated results. To do so, we make thin slices of the Mach number 
distribution like the ones plotted in Fig.~\ref{fig:secuencia_shock}. In these slices, we identify 
the halo  and associate by eye the thin quasi-spherical shell having high and homogenous Mach numbers to 
the shock wave that wraps the halo. Then, considering several directions,  we measure different values for the radial 
distance from the centre of the halo to the shock front.
The final value for the shock radial distance, $R_{s}$, is obtained as the arithmetic mean. 
Using the same procedure, the modules of the gas velocity at the different points of the shock shell used
to estimate $R_{s}$, are computed and averaged to produce the final shock speed, $v_{s}$.

\begin{table}
\begin{center}
\caption{Main data  describing the most massive halo and its 
associated external shock wave. The different columns stand for
the redshift, z, the halo virial mass, $M_{vir}$, in units of 
$10^{14}\,M_\odot$, the halo virial radius, $R_{vir}$, in comoving Mpc, 
the radius of the shock wave, $R_s$, in comoving Mpc, and the shock
speed, $v_s$, in $km/s$, respectively.}
\label{tab1}
\begin{tabular}{ccccc}
\hline 
z & $M_{vir}$ & $R_{vir}$ & $R_{s}$ & $v_{s}$ \\
& $(10^{14}\,M_\odot)$ & $(Mpc)$ &  $(Mpc)$ & (km/s) \\ 
\hline 
 0.0   & 8.3  &   2.4    &    9.5    & 901 \\
 0.13  & 6.1   &  2.1    &   8.4   &    900  \\ 
 0.2  & 2.9   &  1.6    &   7.7   &  920 \\
 0.3  & 2.7   &  1.5    &  6.9   &  900  \\
 0.4 & 2.6   &  1.5    &    6.4   &  970  \\
 0.5  & 2.8   &  1.5    &    5.5   &  950  \\
 0.6 & 2.2   &  1.4    &    4.4  & 945  \\ 
 0.8 & 0.92   &  1.0  &    3.3   &  942  \\   
 1.0 & 0.81   &  0.9  &    2.6   &  950  \\
 1.2 & 0.61  &   0.8  &    2.2   &  900  \\
 1.4 & 0.46   &  0.7  &    1.5   &  890  \\        
\hline
\end{tabular}
\end{center}
\end{table}

\begin{figure}
\centering\includegraphics[width=8.5 cm]{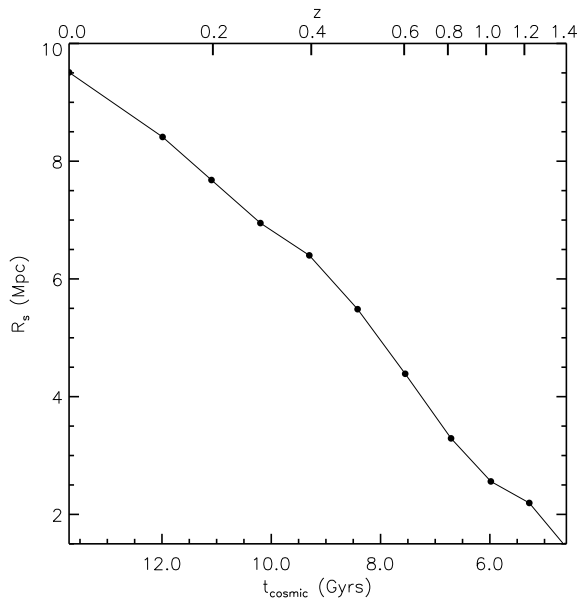}
\caption
 {Temporal evolution of the shock wave distance to the halo centre. }
\label{fig9}
\end{figure}

The propagation of the shock wave surrounding the considered halo can 
be well summarised in Fig.~\ref{fig9}, where the distance from the 
shock wave to the halo centre ($R_s$) is plotted against the time. As expected, the behaviour 
is consistent with the self-similar nature of an spherical isolated shock wave, being the slope of 
the line plotted in Fig.~\ref{fig9}  the propagation speed of the shock. 

The 
last column in  Table~\ref{tab1} shows the  shock speed ($v_s$) at different times.  These 
values are consistent with a constant propagation speed. Nevertheless, there are deviations from 
a perfectly constant slope which  are explained by the errors associated to the graphical method used to 
measure the positions and velocities of the shock wave,  
and by the non idealised environment surrounding the haloes where the external 
medium is far from being homogeneous.

Assuming all these uncertainties, we can use Fig.~\ref{fig9} to date 
the formation time of the studied halo around 
$z\sim 1.4$  and,  therefore, it is nowadays $\sim 8.9 \, Gyrs$ old. 
This redshift is the highest one at which the shock appears perfectly visible and well defined. At early stages, 
the situation is too messy and the crude method used to track the shock produces too much error. In this sense,
this time should be taken as a rough estimate of the collapse of the halo. In any case, it is interesting to compare  
this formation time of the halo with its half mass formation 
time which for this particular cluster is  $z\sim 0.2$.

Under the hypothesis that 
the shocks associated with 
haloes could be observed and measured, 
we suggest the idea to use them as probes to infer 
the main characteristics of the haloes where they were 
created. 
Although the detection 
and measurement of large-scale shock waves is an observational 
challenge, there is hope that new coming instruments 
-- i.e.,  SKA\footnote{http://www.skatelescope.org} 
\citep[Square Kilometre Array; e.g.,][]{rawlings04, batt09} -- 
could shed some light on this scenario.

As it has been already mentioned, 
the properties of the shock waves have been derived by a 
graphical method. Besides, our sample is very reduced. Therefore, due to the uncertainties 
of the method and the statistical limitations, the results of this Section should  be interpreted with caution, 
and only as a way to illustrate the potentiality of the study and observation of the shock waves associated to
cosmic structures. 
 
\section{Summary and conclusions}

In this paper, we focus on the properties of the shock waves
developed during the hierarchical evolution of the cosmic structures by 
means of a  cosmological AMR simulation.

To analyse  the shock waves,  we develop  a numerical algorithm  able of
identifying shocks  in 3-D AMR  simulations.  After labeling all  the shocked
cells   (compression   regions  with   $\nabla\cdot{\bf   v}<0$)  within   the
computational box, the Mach numbers of the shocks are computed by means of the
Rankine-Hugoniot temperature-jump condition.  
The Eulerian nature of the adopted numerical scheme allows to accurately detect 
shocks. Besides, the AMR structure of the employed numerical method produces sets of 
grids mapping different levels of resolution that naturally permits to find 
shocks corresponding to different spatial scales.

We analyse the morphology of the shock patterns detected in the 
simulated volume which turns out to be rather complex. 
In qualitative agreement  with 
previous works \citep[e.g.,][]{ryu03, pfrommer06, vazza09}, 
the shocks morphologies follow the shape of 
the cosmic web, being 
filaments, sheets, and haloes surrounded by strong external shocks, 
while more irregular and weaker internal shocks are found within 
the virial radius of haloes. From the morphological point of view, 
it is remarkable that the external shocks associated to the most 
massive haloes -- galaxy clusters -- show a quasi-spherical shape
around them.

The statistical analysis of the shocks found in the simulation has 
produced several important conclusions. We have estimated that the 
filling factor of shocks -- the fraction of the total volume 
occupied by shocked cells -- at  $z=0$ is roughly the $20 \%$ of 
the simulated  volume   with a mean Mach number of $\simeq 4$.

Following with this analysis, we have computed the 
differential shock distribution function 
within the simulated volume  as a function of  the shock Mach
numbers. This function can be fitted  by   two    different    power   
laws    with   the   form
$dN(\mathcal{M})/d\mathcal{M}  \propto \mathcal{M}^{\alpha}$, 
having the  low Mach
numbers  (up to  $\simeq 20$)   a slope  of $\alpha  \simeq -1.7$,
whereas  a steeper  relation ($\alpha  \simeq -4.1$) stands for stronger
shocks.  
This turn has a crucial meaning as it shows the  
transition between  the scales  associated to
internal and external shocks. The distribution function also 
reveals  that  most part of cosmological shocks are essentially
weak  shocks  ($\mathcal{M}  \leq  2$), as already demonstrated 
by other authors \citep[e.g.,][]{ryu03, pfrommer06, vazza09}.

In  agreement with the information derived from the shock 
distribution function, we find  that the average Mach  
number within  the virial
radius  of  haloes at  $z=0$  is $\mathcal{M}  \approx  5$.
If we look at higher redshifts, this  average Mach  
number  is always below $\approx 20-25$, which perfectly
correlates with the turn obtained in the shock  distribution function.

When haloes, according to their properties,  are located in the 
plane  formed by the two axes, virial mass  and average 
Mach number inside the virial radius,  
 we observe two well-separated trends at $z=0$. 
One stands for a group of haloes with almost 
constant low Mach numbers (up to 5) 
which  seems to be  non dependent on  the halo mass. 
This population of haloes, characterised by low virial Mach
numbers, could stem  from the primordial  formation of 
structures which have been initially set up in a  
quasi-relaxed and smooth process.  

The second group of haloes spreads  all over the range of Mach numbers 
but showing a clear correlation with the halo masses.
They  represent the haloes where 
strong shocks took place during
their early evolution as a  consequence of early mergers and
violent accretion processes.   We conclude that  the 
evolution  of  haloes  along  the plane  $\mathcal{M}-M_{vir}$  is  intimately
related with their merging history and with their initial set up. 
In that sense, possible  estimates of 
$\mathcal{M}$ for a given halo could lead to know whether it has suffered 
merger events in the past, and it may correlate with other relevant 
features like the existence or not of cool cores.

Finally, we speculate about the possibility that new radio observations
may observe the shocks associated with the formation and evolution of the 
cosmic structures. Assuming that this could be feasible in the near future,
we propose to turn the argument around, and to use the main features of shocks
 -- position and strength -- to infer import features of the halo harbouring 
the shock like, for instance, the formation time of the halo, or its total mass.
This last application could  be a powerful new
tool  for cluster  mass estimation  if we  are able  to detect  the accretion
shocks with radio observations \citep[e.g.,][]{giacintucci08}.

Our main conclusion is that shock waves play a crucial role in the 
formation and evolution of galaxies and galaxy clusters. 
Despite this general and obvious statement, 
direct evidence of shocks, 
both from the cosmic web formation processes (large scales)  
and those due to cluster merging
events (small scales),  has  been found  only  in  a relatively
 small  number  of  clusters thanks  to
observations of radio  relics and temperature maps in  X-rays.  
Therefore, we believe that it is necessary to pursue the study of the role 
of shocks in cosmological context  as they are fundamental players 
in the paradigm of structure formation in the Universe. 
Their complete theoretical description together with  their detection and observation 
are still a challenge.

\section*{Acknowledgements}
The authors would like to thank Stefano Borgani for valuable discussions and 
the referee for her constructive criticism.
This work has  been supported by {\it Spanish  Ministerio de Ciencia e
Innovaci\'on}     (MICINN)     (grants    AYA2010-21322-C03-02     and
CONSOLIDER2007-00050) and Generalitat Valenciana 
(grant PROMETEO/2009/103).   
SP acknowledges a fellowship from the European Commission's Framework 
Programme 7, through the Marie Curie Initial Training Network CosmoComp 
(PITN-GA-2009-238356).
Simulations  were   carried  out   in  the   {\it  Servei
d'Inform\'atica de  la Universitat de Val\`encia} 
using the {\it Lluis Vives} supercomputer.

\end{document}